%Paper: astro-ph/9412044
%From: Liliya Rodrigues <llrw@bluemoon.astro.washington.edu>
%Date: Tue, 13 Dec 94 14:05:26 -0800

\magnification=\magstep1
\hsize=6.0truein
\vsize=8.75truein
\baselineskip=12.045truept % sets baseline spacing to 6 lines per inch %
\parindent=2.5em     % sets paragraph indent to 5 spaces %
\parskip=0pt
\def\hi{\noindent \hangindent=2.5em}
\def\bigskip{\vskip 12.045pt}  % skip one line %
\overfullrule=0pt
\font\twelverm=cmr10 at 12truept  % sets type size to 12 pt %
\twelverm    % document font 12 point Roman %
\nopagenumbers
\def\title#1{\centerline{\bf #1}}
\def\author#1{\bigskip\centerline{#1}}
\def\address#1{\centerline{#1}}
\def\sec#1{\bigskip\centerline{#1}\bigskip}
%%%%%%%%%%%%%%%%%%%%%%%%%%%%%%%%%%%%%%%%%%%%%%%%%%%%%%%%%%%%%%%%%%%%%%%%%%%%

\def\sqdeg{$\sqcup\mskip-12.0mu \sqcap^\circ$ }
%%%%%%%%%%%%%%%%%%%%%%%%%%%%%%%%%%%%%%%%%%%%%%%%%%%%%%%%%%%%%%%%%%%%%%%%%%%%

\def\go{
\mathrel{\raise.3ex\hbox{$>$}\mkern-14mu\lower0.6ex\hbox{$\sim$}}
}
\def\lo{
\mathrel{\raise.3ex\hbox{$<$}\mkern-14mu\lower0.6ex\hbox{$\sim$}}
}
%%%%%%%%%%%%%%%%%%%%%%%%%%%%%%%%%%%%%%%%%%%%%%%%%%%%%%%%%%%%%%%%%%%%%%%%%%%%

{}~~
\bigskip
\title{GRAVITATIONAL LENSING OF VARIABILITY SELECTED}
\title{QSOs BY GALAXY CLUSTERS }
\author{L. L. Rodrigues-Williams}
\address{Astronomy Dept, FM-20, Univ. of Washington, Seattle WA 98195, USA}
\address{llrw@astro.washington.edu}
\author{M. R. S. Hawkins}
\address{The Royal Observatory, Blackford Hill, Edinburgh EH9, UK}
\address{mrsh@starlink.roe.ac.uk}

\sec{ABSTRACT}

   Distribution and properties of QSOs behind galaxy clusters in
the UKJ287 field were studied. QSOs were selected using variability
criteria and are confined to $z \ge 0.4$ and $m_B \le 19.5$.
No reddening or obscuration of background QSOs due to dust in
clusters is detected.
   A statistically significant positive angular correlation between
clusters ($\langle z\rangle \simeq 0.15$) and QSOs ($0.4 \le z \le 2.2$)
on scales of several arcminutes is found. This association of QSOs with
foreground clusters is ascribed to gravitational lensing by galaxy
clusters. The amplitude of associations displays trends expected if
weak lensing by clusters were responsible:
(a) most statistically significant associations are found for
brightest QSOs found very close in projection to clusters centers;
(b) QSO number excess behind clusters increases closer to cluster
centers; and (c) QSO number excess increases with brighter QSO flux
limit in a way that would be expected from a simple lensing model
derived using Boyle, Shanks \& Peterson (1988) QSO number counts
and an average amplification of $A=2$ due to clusters. The implied
amplification is substantially larger than would be expected from
isothermal sphere clusters with a velocity dispersion
$\sigma_v \sim 1000$ km/s.

\sec{1. QSO NUMBER COUNTS}

Since UKJ287 QSOs were selected using variability criteria
(see Hawkins \& Veron 1993, HV for details) as opposed
to a more conventional color criteria it will be
instructive to compare the HV and BSP (Boyle, Shanks \& Peterson, 1988)
QSO number-magnitude counts. The latter were UVX selected.

The HV QSO number counts are shallower than BSP at the faint end, around
$m\simeq 19$. Assuming magnitude calibration is the same in both
samples QSO number density at $m=19.5$ is 5.6 and 8.4/\sqdeg for
HV and BSP respectively. However
using the Kolmogorov-Smirnov test we cannot rule out the hypothesis that
HV and BSP QSOs were drawn the same distribution. For the
total numbers of HV and BSP QSOs down to $m_{lim}=19.5$---106 and
about 97 QSOs respectively---the KS test confidence level is only 67\%.

\sec{2. QSO-CLUSTER CORRELATION}

Galaxy cluster positions were cross-correlated with those of background
QSOs ($z \ge 0.4$). In order to take into account different angular
sizes of clusters angular separations between QSOs and cluster centers
were calculated in terms of individual cluster radii,
then QSO-cluster separations were multiplied by the
average cluster radius ($0.06^\circ$) to yield degrees.
Fig.1 shows two-point correlation functions for four different QSO
flux cutoffs, $m_{lim}$=18.0, 18.5, 19.0, and 19.5. Each point
in Fig.1 represents a separation of one average cluster radius.

Clusters and (at least the bright) QSOs are seen to be positively
correlated.
How significant are these correlations and can they be attributed to
gravitational lensing? We will explore these two questions together.
We make no attempt in this paper to investigate alternative explanations
of foreground cluster/background QSO associations (but see
Rodrigues-Williams \& Hogan 1994).

A useful way of evaluating statistical significance of such observations
is with binomial statistics (see Seitz \& Schneider 1994).
We calculated binomial probability of having the observed number of
QSOs within 0.5, 1.0, 1.5, 2.0, and 2.5 cluster radii with different
QSO flux limits.
The most statistically significant associations occur for QSOs within
0.5 cluster radii and with $m_{lim}=18.0,18.1,$ and 18.2 (99.3\%,
99.3\%, and 98.6\% c.l. respectively), and within 2.0 cluster radii
and $m_{lim}=18.8$ (98.6\% c.l.). While these probabilities may not be
overwhelmingly convincing by themselves there is other evidence that
QSO-cluster associations are not chance associations but
are due to weak lensing by clusters: (a) the most statistically
significant results occurs for the brightest QSOs found closest
in projection to cluster centers; (b) QSO number excess behind clusters
increases closer to cluster centers, see Fig.1; (c) QSO number
excess increases with brighter QSO flux limit, see Fig.1 and 2.

Thus we conclude that there is a statistically significant
association between QSOs and clusters in our sample and the association
is due to gravitational lensing.

\sec{3. GRAVITATIONAL LENSING}

With our limited number of QSOs we cannot afford to
investigate QSO overdensity as a function of distance form cluster
centers. Instead,
let us ``fix" cluster radii at twice their cataloged values, and
thus divide QSOs into ``association" QSOs, $i.e.$ those found behind
extended galaxy clusters, and ``field", $i.e.$ the rest of the QSOs.
With this division association area is 17.4\% of the total UKJ287 field
area. Let us define overdensity, $q$, as the ratio of the actual number
of association QSOs to the number that would have been found had
the field QSO number density applied to the area behind clusters.

Filled dots in Fig.2 are the observed overdensities of QSOs behind
extended clusters.  Error bars represent r.m.s. dispersion between
clusters. We will now compare these observations with what might be
expected from a simple model. Assume the clusters are uniform circular
disks with constant amplification, $A$, at an average redshift of 0.15,
while QSOs are all located at large redshifts and obey BSP
magnitude-number counts. This model is represented by three solid lines
in Fig.2a, each for a different value of $A$: 1.14, 2.0, and 3.0.
$A$=2 model seems to fit the observations reasonably well.

The dashed lines in Fig.2a were obtained using the modified BSP counts:
dashed lines to the left of the solid lines are the same as BSP model,
$except$ the break in the counts has been moved brightward by 0.5
mag. to $m_{break}$=18.65. Similarly, the dashed lines on the right
have $m_{break}$=19.65, or 0.5 mag. fainter that in the BSP counts.
The $q$ vs. $B_{lim}$ plot is sensitive to the $m_{break}$ location, in
particular, modified BSP counts with $m_{break}$=19.65 are not consistent
with the observations regardless of the value of $A$.

Solid lines in Fig.2b use the number counts fit to the observed
HV counts. Again, three amplifications are represented by three
solid lines: $A$=1.14, 2.0, and 6.0. $A$=6 clusters could reproduce the
observations, however as we will see below such large amplifications
are not characteristic of galaxy clusters.

What kind of clusters would produce observed overdensities?
A back-of-an-envelope calculation shows that most clusters should
not be able to generate the observed QSO overdensities.
With $\langle z_{cluster}\rangle\simeq 0.15$ and
$\langle\theta_{cluster}\rangle\simeq 0.12^\circ$
average extended cluster radius is $R_{cluster}\simeq 0.75h^{-1} Mpc$.
$A=2$ implies average surface mass density of
clusters, $\sigma\simeq 0.3$ in terms of critical surface mass density
for lensing. The latter is about 1 $gm/cm^2$ for lenses at
$\langle z_{cluster}\rangle\simeq 0.15$ and sources at
$\langle z_{QSO}\rangle\simeq 1.5$.
Assuming a singular isothermal sphere model for clusters, their velocity
dispersion is calculated to be $\sigma_v\simeq 2130 km/s$. This value is
too high compared to observations of cluster velocity dispersions,
hence clusters appear to be producing larger overdensities than expected.

What amplification would reasonable clusters be expected to produce?
Using the same model as above, clusters with $\sigma_v\sim 1000 km/s$
will have A=1.14. The overdensity curves for this value of A are shown
in Fig.2a,b. In both cases these are not consistent with the observations.

\sec{4.  DUST IN THE UKJ287 CLUSTERS}

Dust in clusters will affect background QSOs in two ways:
(1) obscuration will diminish the number density
of QSO behind clusters, and (2) QSOs behind clusters would
be reddened. In a sample of variability selected QSOs one would not
expect to ``lose" QSOs to reddening.
Since $A_B\approx 4\times E(U-B)$, obscuration
is expected to be the more important of the two dust effects.

(1) Obscuration: An overdensity of QSO behind clusters is observed.
Underdensity is not observed at any
QSO limiting magnitudes (see Fig.1), hence there is very little,
if any obscuration due to dust in clusters.

(2) Reddening: All QSOs with $z \ge 0.4$ were divided into two samples:
those behind extended clusters and the rest.
The $(U-B)$ color distribution of association and field QSOs is not
consistent with dust reddening by clusters. In fact,
the average color of association QSOs is somewhat bluer than field QSOs.

We conclude that there is little or no smoothly distributed dust in
UKJ287 galaxy clusters.

\sec{5. CONCLUSIONS}

Statistically significant association between high redshift
variability selected QSOs and foreground galaxy clusters were
detected. The amplitude of associations displays trends expected if
weak gravitational lensing by clusters were responsible:
(a) most statistically significant associations are found for
brightest QSOs found very close in projection to clusters centers;
(b) QSO number excess behind clusters increases closer to cluster
centers (Fig.1); and
(c) QSO number excess increases with brighter QSO flux limit (Fig.1,2).

\sec{REFERENCES}

\hi{Boyle, B. J., Shanks, T., \& Peterson, B. A. 1988, MNRAS, 235, 935, BSP}

\hi{Hawkins, M. R. S., \& Veron, P. 1993, MNRAS 260, 202, HV}

\hi{Rodrigues-Williams, L. L., \& Hogan, C. J. 1994 AJ, 107, 451}

\hi{Seitz, S., \& Schneider, P. 1994, Preprint}

\vfill
\bye